\def\PRL{ Phys. Rev. Lett.}
\def\PRD{ Phys. Rev. D}
\def\NPB{ Nucl. Phys.}
\def\PLB{Phys. Lett. B}
\def\vev#1{\langle #1 \rangle}
\def\be{\begin{equation}}
\def\ee{\end{equation}}   
\def\bq{\begin{eqnarray}}
\def\eq{\end{eqnarray}}
\def\dld{\Delta_L^\dagger}
\def\dl{\Delta_L}
\def\drd{\Delta_R^\dagger}
\def\dr{\Delta_R}
\def\p{\phi}
\def\pd{\phi^\dagger}
\documentstyle[12pt]{article}
\font\tenrm=cmr10
\title{\huge Topological Defects in the Left-Right Symmetric Model and 
their Relevance to Cosmology}

\author{U. A. Yajnik$^a$\thanks{yajnik@niharika.phy.iitb.ernet.in} ,
 Hatem Widyan$^b$ \thanks{hatem@ducos.ernet.in} , D. Choudhury$^c$
\thanks {debchou@mri.ernet.in} ,\\ S. Mahajan$^b$
\thanks{sm@ducos.ernet.in} , A. Mukherjee$^b$
\thanks{am@ducos.ernet.in} \\ 
{$^a${\tenrm Physics Department, Indian Institute of Technology
Bombay, Mumbai 400\thinspace076, India}} \\
{$^b${\tenrm Department of Physics and Astrophysics, University of
Delhi, Delhi 110 007, India}} \\
{$^c${\tenrm Mehta Research Institute of Mathematics and Mathematical  
	Physics}}\\ {\tenrm Chhatnag Road, Jhusi, Allahabad 221 506, India}}

\date{}
\begin{document}
\maketitle
\baselineskip=0.7cm
\begin{abstract}
It is shown that the minimal left-right symmetric model admits
cosmic string and domain wall solutions. 
The cosmic strings arise when the $SU(2)_R$ is broken and can either be 
destabilized at the electroweak scale or remain stable through the 
subsequent breakdown to $U(1)_{EM}$.
The strings carry zero modes of the neutrino fields. Two distinct
domain wall configurations  exist above the electroweak phase
transition and disappear after that. Their destabilization provides
new sources of non-equilibrium effects below the electroweak scale
which is relevant to baryogenesis.
\end{abstract}
%
%
\begin {section}*{I. INTRODUCTION}
Topological defects are regions of trapped energy density which can
be produced at the time of cosmological phase transitions and
survive after that if the topology of the
vacuum manifold of the theory is nontrivial. Typically, cosmological
phase transitions occur when a gauge symmetry of a particle physics
theory is spontaneously broken. In that case, the cores of the
topological defects formed are regions in which the symmetry of the
unbroken theory is restored. The defect formation and stability
conditions are as follows \cite{Kibble}. Consider the spontaneous
symmetry breaking of a group $G$ down to a subgroup $H$ of
$G$. Topological defects, arising according to the Kibble mechanism
\cite{Kibble} when $G$ breaks down to $H$, are classified in terms of
the homotopy groups of the vacuum manifold ${G}/{H}$
\cite{Kibble,Coleman}. The relevant homotopy groups are $\Pi_i(G/H),~
i=0,1,2$. If $\Pi_i(G/H)$ is nontrivial, topological defects can
form. For $i=0, 1$ and $2$, the defects are domain walls, cosmic
strings and monopoles respectively. We are typically interested in a
scenario where $H$ breaks further to $K$. If $\Pi_i(G/K)$ is
nontrivial, defects are possible in this second stage of symmetry
breakdown. Thus, if $\Pi_i(G/H)$ and $\Pi_i(G/K)$ (for some $i$) are
both nontrivial, the defect formed in the first stage persists in the
second stage. If, on the other hand, $\Pi_i(G/K)$ is trivial, then the
corresponding defect does not exist in the second stage. Thus, the
defects formed in the breaking of $G$ to $H$ must be unstable when $H$ 
breaks to $K$. Cosmic strings can explain large
scale structure, anisotropies in cosmic microwave background radiation
(CMBR), and part of the baryon asymmetry of the universe
\cite{kib,sbdanduay1,rbetal1,garvacbar,rbetal2}. Global monopoles can
explain structure formation in the universe. Domain walls and
local monopoles, on the other hand, if they exist, are potentially
problematic. They would dominate the energy density of the universe
and overclose it \cite{kib,vachasrev,Preskill,vilenshel}. 

 The problem of monopoles is especially
serious since it is generic to grand unification scenarios 
\cite{Preskill}. The popular solution based on the idea of inflation
cannot be implemented in the minimal grand unified theories
(GUTs). Similarly, to solve the domain wall problem \cite{Zeldovich},
we require inflation to take place after the phase transition 
that causes the production of these defects. This is difficult to
achieve in general. 

Recently, a possible solution of the monopole
problem was suggested \cite{Dvali}, based on the possibility that
unstable domain walls sweep away the monopoles. The idea of symmetry
nonrestoration at high temperature $T$
\cite{Weinberg,Senjanovic1,Senjanovic2} provides a simple way out of
the domain wall problem
\cite{Dvali1,Dvali2}. Unfortunately, in case of the monopole problem, 
the situation is far from clear \cite{Dvali3}. In a recent paper, 
Bajic et al \cite{Bajic} show that the monopole
problem in grand unified theories as well as the domain wall problem
may be easily solved if the lepton number asymmetry in the universe is 
large enough. In spite of the fact that domain walls are
undesirable objects, during their decay they can provide a
departure from thermal equilibrium which is one of the conditions for
baryogenesis \cite{Sakharov,Brand,Brand1}.

Currently, several unification schemes are being investigated in
detail, specially for their signatures in the 
planned particle accelerators. Some of the unification schemes have
interesting consequences for cosmology. A rich variety of cosmic 
string solutions was demonstrated \cite{asanduay,uayexcon} in the 
context of $SO(10)$ unification and has received fresh attention
\cite{acdetal}. Furthermore, as the non-viability of several models 
for electroweak baryogenesis is becoming apparent
\cite{shaphiggsbd,carwag,riotto,clijoka}, it is interesting to search
for new mechanisms for low-energy baryogenesis in other unified
models \cite{rbetal2,rj1}.

 As a particle physics model, we consider one of the most attractive
extensions of the standard electroweak model, based on the gauge group
$SU(2)_L \otimes SU(2)_R \otimes U(1)_{B-L}$
\cite{Senjanovic1,Pati}. Various models employing this gauge
group are possible, depending on which Higgs and fermion spectrum is
chosen, and whether or not exact discrete left-right symmetry is
imposed. We are interested in the class of left-right symmetric models
described in \cite{Senjanovic2,Pati,mohapbook}. Besides explaining the
observed parity violation of weak interactions  
at low energies, these models also provide an explanation for the
lightness of ordinary neutrinos, via the see-saw mechanism. 

 In this paper we investigate the minimal left-right symmetric 
model for the presence of topological defect solutions. 
We begin with the phase in which only the first stage of symmetry
breaking $ SU(2)_R \otimes U(1)_{B-L} \to U(1)_Y$ has occurred. 
We show that the cosmic string solution exists
in the high temperature phase of the theory where the
electroweak symmetry is restored. These string defects may either be
destabilized at the electroweak phase transition or may
acquire additional condensates and continue to enjoy
topological stability.  We show that the strings possess zero-energy
modes of the right handed neutrino, and below the electroweak scale,
also those of the left handed neutrino.
The model also admits at least two kind of domain wall solutions which 
are stable only above the electroweak scale.

 In Sec. II, we describe the minimal left-right symmetric model. In
Sec. III, we discuss the possibility of producing cosmic 
strings and associated
zero modes. In Sec. IV, we discuss the domain wall solutions. 
Finally, Sec. V contains the conclusions and the cosmological
consequences of the defects.
\end {section}
%
%
\begin {section}*{II. LEFT-RIGHT SYMMETRIC MODEL}
 We consider the minimal $SU(2)_L \otimes SU(2)_R \otimes U(1)_{B-L}$ 
model with a discrete left-right symmetry
\cite{Senjanovic2,mohapbook,Deshpande}. This model is formulated so
that parity is a spontaneously broken symmetry: the Lagrangian is
left-right symmetric but the vacuum is not invariant under the parity
transformation. Thus the observed V-A structure of the weak
interactions is only a low energy phenomenon, which should disappear
when one reaches energies of order $v_R$, where $v_R$ is the vacuum
expectation value of one of the Higgs fields.

 According to left-right symmetric requirement, quarks ($q$) and
leptons ($\psi$)are placed in left and right doublets,
\bq
q_{L} = {\pmatrix{ u \cr d }}_L \equiv
\pmatrix{\frac{1}{2},0,\frac{1}{3}}, ~ ~ 
q_{R} = {\pmatrix{ u \cr d }}_R \equiv
\pmatrix{0,\frac{1}{2},\frac{1}{3}},\nonumber \\ [0.3cm]
\psi_{L} = {\pmatrix{ \nu_e \cr e }}_L \equiv
\pmatrix{\frac{1}{2},0,-1}, ~ ~
\psi_{R} = {\pmatrix{ \nu_e \cr e }}_R \equiv
\pmatrix{0,\frac{1}{2},-1} ~ ~ ,
\eq
where the representation content with respect to the gauge group is
explicitly given. Since the weak interactions observed at low
energies involve only the left handed helicity components, the
electric charge formula can be written in a left-right symmetric form
as 
\be
Q=T^3_{L} + T^3_{R} + \frac{B-L}{2} ~~ ,
\ee
where $T^3_L$ and $T^3_{R}$ are the weak isospin represented by
$\tau^3/2$, where $\tau^3$ is the Pauli matrix.
Regarding the bosons, gauge vector bosons consist of two tripltes
$W_L^\mu \equiv (3,1,0)$, $W_R^\mu \equiv (1,3,0)$ and a singlet
$B^\mu \equiv (1,1,0)$. 

 The Higgs sector of the model is dictated by two requirements, the
choice of the symmetry breaking term and the desire to reproduce the
phenomenologically observed light masses of the known neutrinos via
the see-saw mechanism. Then the unique minimal set is
\begin{eqnarray}
\Phi = \pmatrix{\phi_1^0 & \phi_1^+ \cr \phi_2^- & \phi_2^0} \equiv
\pmatrix{\frac{1}{2},\frac{1}{2},0} ~~, \nonumber \\
\Delta_L = \pmatrix{\frac{\delta_L^+}{\sqrt{2}} & \delta_L^{++} \cr
\delta_L^0 & -\frac{\delta_L^+}{\sqrt{2}}} \equiv (1,0,2) ~~ ,
\nonumber \\
\Delta_R = \pmatrix{\frac{\delta_R^+}{\sqrt{2}} & \delta_R^{++} \cr
\delta_R^0 & -\frac{\delta_R^+}{\sqrt{2}}} \equiv (0,1,2) ~~.
 \label{Higgs} 
\end{eqnarray}
where the scalar fields have been written in a convenient
representation using  $2 \times 2$ matrices.
 
 The potential energy of the Higgs fields cannot have trilinear terms, 
This can be seen as follows. Since the triplets $\Delta_L$ and 
$\Delta_R$ have nonzero $B-L$, these must always appear in the
quadratic combinations $\Delta_L^\dagger \Delta_L$, $\Delta_R^\dagger 
\Delta_R$, $\Delta_L^\dagger \Delta_R$ or $\Delta_R^\dagger
\Delta_L$. These can never be combined with a single bidoublet $\Phi$ 
in such a way as to form $SU(2)_L$ and $SU(2)_R$ singlets. However,
quartic combinations of the form $Tr(\Delta_L^\dagger \Phi \Delta_R
\Phi^\dagger)$ are in general allowed by the left-right
symmetry. According to these strict conditions, the most general form 
of the Higgs potential is (see \cite{mohapbook})
\be
V = V_\Phi + V_\Delta + V_{\Phi \Delta} ~~ , \label{potential}
\ee
with
\bq
V_\Phi  = -\sum_{i,j} \mu^2_{ij} Tr \Big(\pd_i \p_j\Big) & + &
\sum_{i,j,k,l} \lambda_{ijkl} Tr\Big(\pd_i \p_j\Big) Tr\Big(\pd_k
\p_l\Big)  
\nonumber \\ [0.3cm]
& + &
\sum_{i,j,k,l} \lambda^\prime_{ijkl} Tr\Big(\pd_i \p_j \pd_k \p_l\Big)
~~ .  
\label{vphi}
\eq
\bq
V_\Delta  = & - & \mu^2 Tr\Big(\dld \dl + \drd \dr\Big) 
 + 
\rho_1 \Big[\Big(Tr(\dld \dl)\Big)^2 + \Big(Tr(\drd \dr)\Big)^2\Big]
 \nonumber \\ [0.3cm] 
& + & 
\rho_2 \Big[Tr\Big(\dld \dl \dld \dl\Big) + Tr\Big(\drd \dr \drd
\dr\Big)\Big]  
+ 
\rho_3 \Big[Tr\Big(\dld \dl\Big) Tr \Big(\drd \dr\Big)\Big] \nonumber
\\ [0.3cm] 
& + &
\rho_4 \Big[Tr\Big(\dld \dld\Big) Tr\Big(\dl \dl\Big) 
 + 
Tr\Big(\drd \drd\Big) Tr\Big(\dr \dr\Big)\Big]   ~~ . \label{vdelta} 
\eq
\bq
V_{\Phi\Delta} & = & \sum_{i,j} \alpha_{ij} Tr\Big(\pd_i \p_j\Big)
Tr\Big(\dld \dl + \drd \dr\Big) \nonumber \\ [0.3cm]
& + &
\sum_{i,j} \beta_{ij} Tr \Big[\p_i \pd_j \dld \dl + \pd_i \p_j \drd
\dr\Big] 
\nonumber \\ [0.3cm]
& + &
\sum_{i,j} \Big[ \gamma_{ij} Tr\Big(\dld \p_i \dr \pd_j\Big)+
h.c.\Big] ~~ . 
\label{vdeltaphi} 
\eq
 Note that, as a consequence of the discrete left-right symmetry, all
terms in the potential are self-conjugate, except for $\alpha_{ij}$;
therefore $\alpha_{ij}$ is the only parameter which may be
complex. Since we will not discuss the CP violation aspect of the
generation of baryon asymmetry, we assume $\alpha_{ij}$ to
be real. It can be shown \cite{Deshpande} that, without fine tuning,
$\gamma_{ij}$ terms spoil the seesaw mechanism by inducing a direct
Majorana mass term for the left-handed neutrino. Therefore, we set
$\gamma_{ij}=0$ in our calculations. This choice will also avoid the
unwanted presence of  large flavor changing neutral current (FCNC).

 Moreover, only the neutral components of the scalar fields,
$\phi_1^0$, $\phi_2^0$, $\delta_L^0$, $\delta_R^0$, can acquire vacuum 
values (vevs) without violating electric charge. If $\Delta_L$ or
$\Delta_R$ acquire a vacuum expectation value (vev), then $B-L$ is
necessarily broken, and if $<\Delta_L> \neq <\Delta_R>$, parity
breakdown is also ensured. Thus the following vevs are sufficient for
achieving the correct pattern of symmetry breaking  
\be
\Phi = \pmatrix{\kappa & 0 \cr  0 & \tilde\kappa}, ~~
\Delta_{L,R} = \pmatrix{ 0 & 0 \cr v_{L,R} & 0} ~~, \label{vevs}
\ee
where $\kappa$, $\tilde\kappa$, $v_L$ and $v_R$ are taken to be real, 
and phenomenologically the hierarchy $ \kappa \ll v_R$, $v_L \ll
\tilde\kappa $ is required. 

Fermion masses are obtained from Yukawa couplings of  quarks and 
leptons with the Higgs bosons. For one generation of quarks $q$ and
leptons $\psi$, the couplings are given by \cite{mohapbook}
\begin{eqnarray}
{\mathcal L}_Y & =  & h^q \, \bar q_L \,  \phi \, q_R + \tilde h^q \,
\bar q_L \, \tilde\phi \, q_R   \nonumber \\
 & + & h^l \, \bar\psi_L\,  \phi \, \psi_R + \tilde h^l \, \bar\psi_L
 \, \tilde\phi \,  \psi_R  \nonumber \\
& + & f \left( \psi_L^T \, C^{-1} \, \tau_2 \, \Delta_L \, \psi_L +
 \psi_R^T \,  C^{-1} \, \tau_2 \, \Delta_R \, \psi_R \right) +
 h.c. ~~.\label{yukawa} 
\end{eqnarray}
The Majorana mass terms allowed for the neutrinos are a source of
lepton number violation as well as CP violation.
The couplings are also important for studying
fermionic zero-modes of cosmic strings.

\end{section}
%
%
\begin {section}*{III. COSMIC STRINGS AND FERMION ZERO-MODES} 

 In this section we discuss the cosmic string sectors occurring in this
theory. In the following we use the notation  $X=(1/2)(B-L)$. 
Consider  first a pure $SU(2)$ theory with a two real triplet
scalars which break the symmetry. 
A cosmic string sector exists in
this breakdown because the stability group of the vev is $Z_2$. The
$Z_2$ arises because the $SU(2)$ element $-I$, negative of the
identity, leaves invariant 
the vev of the triplets \cite{vilenshel}. 

Consider next the breakdown of $SU(2)_R \times
U(1)_{B-L}$ due to $\Delta_R$. The scalar field $\Delta_R$ is complex
and  can be parametrized as 
\be
\Delta_R = (\vec r + i \vec s). \vec \tau ~~,\label{deltaparam}
\ee
where $\vec r$ and $\vec s$ are  3-dimensional real vectors and
$\tau_a$ are the Pauli matrices. The vev for $\Delta_R$ in
Eq.~(\ref{vevs}) implies that $\langle \vec r \rangle = (1,0,0)$ and
$\langle \vec s \rangle = (0,-1,0)$, in the usual 
3-dimensional basis. If $SU(2)_R$ alone were present, a cosmic string
sector would exist in this breakdown as discussed above. Fortunately,
the inclusion 
of $U(1)_{B-L}$ does not change this conclusion. To show this, suppose
the cosmic string ansatz is set up as usual by a path in $SU(2)_R$
connecting $I$ to $-I$ by $2\pi$ rotation generated by a broken
generator. We may 
try to unwind this using the surviving gauge symmetry $U(1)_Y$. But a
$2\pi$ rotation generated by $Y=T_R^3 + X$ also leads to a $2\pi$
winding in the $U(1)_{B-L}$ space. Thus the unshrinkable path
persists. This reasoning also shows that the $Z$ expected from the
breakdown of $U(1)_{B-L}$ group by itself does not persist due to the
presence of the $SU(2)_R$. Rotation by the unbroken generator $Y$
identifies distinct sectors labelled 
by $Z$ modulo the $Z_2$ which survives the $SU(2)_R$ breakdown
\cite{footnote}. 

Consider next, the breakdown of the model to electromagnetism. If the
only additional field were $\Delta_L$, the residual symmetry would be
$Z_2 \times Z_2$ by a simple extension of the previous
arguments. Specifically, the $Z_2 \times Z_2$ elements are $(I,I)$,
$(-I,I)$, $(I,-I)$, $(-I,-I)$ in an obvious notation. However, the vev
of the bidoublet $\Phi$ (Eq.~(\ref{vevs})) is invariant only under
$(I,I)$, $(-I,-I)$. The $Z_2$ consisting of these two elements is
therefore a discrete symmetry of the low temperature
theory. In the following we set up ansatze for the cosmic
strings both in the high temperature and low temperature phases
exploiting the $Z_2$ relevant to each phase.

 Let an infinite long string be oriented along the
$z$ axis and let $\theta$ be the angle in the $x$-$y$ plane. We
construct a map $U^\infty(\theta)$ from the infinitely large circle,
$(S^1)^\infty$, in the $x$-$y$ plane into some one-parameter $U(1)$
subgroup of the parent group generated by a broken generator
$K$. Consider first the high temperature phase. Since  $T_R^3$ and $X$
are the diagonal generators of the parent group and $Y$ is preserved,
the orthogonal combination $\tilde Y = T_R^3 - X$ is a good choice for
$K$. Consider the general map given by
\be
U^\infty(\theta;p) =  e^{ip \tilde Y \theta} = 
\pmatrix{e^{ip\theta/2} & 0 \cr 0 & e^{-ip\theta/2}} \circ  
e^{-ip \theta X} ~~,\label{uinfp} 
\ee
where $p$ is a real parameter to be determined.
The notation $\circ$ is to keep distinct the $U(1)_{B-L}$ which is
multiplicative from the $SU(2)_R$ whose action is adjoint.
Since $SU(2)_R$ acts on $\Delta_R$ by similarity transformation, the
resulting general scalar field ansatz is
\be
\Delta_R(\infty,\theta;p)= \pmatrix{ 0 & 0 \cr v_R 
e^{-i2p \theta} & 0} ~~.\label{deltarp}
\ee
The minimal value of $|p|$ required to make the ansatz single-valued is
$1/2$. Notice that both the values $\pm 1/2$ of $p$ belong to the same
topological sector. For $p=+1/2$, a rotation by $\pi$ generated by $Y$
deforms the 
path to be the entirely in $SU(2)_R$, connecting $I$ to $-I$. For
$p=-1/2$, the same is done by a $Y$ rotation by $-\pi$. For $p=1$,
$U^\infty(\theta)$ winds once around $U(1)_{B-L}$ and is a $2\pi$
rotation in $SU(2)_R$. This path can be deformed by $U(1)_Y$ to be a
purely $4\pi$ rotation in $SU(2)_R$ and thus it is trivial. By
extending this reasoning to the values $p=\pm,1, \pm2, \cdots$, all
such maps can be reduced to the trivial sector. Similarly, all the
paths with $p=\pm3/2, \pm 5/2, \cdots$ can be reduced either to the
$p=+1/2$ or $p=-1/2$ path. Finally, $p=+1/2$ is distinguished from
$p=-1/2$ only by the sense of winding. Rotation by $\pi$ about any
axis in the $x-y$ plane makes them physically indistinguishable. 

The ansatz for $p=1/2$ can
be extended to finite values of the radial coordinate $r$ as follows
\be
~\Delta_R(r,\theta)= \pmatrix{ 0 & 0
 \cr v_R e^{-i  \theta} & 0} f_R(r)~~,
\ee
where $f_R(r)$ is a real function of $r$ satisfying $f_R(0)=0$ and
$f_R(r) \to 1 $ as $ r \to \infty$. 
This completes the ansatz for the scalar field.   

The ansatze for the gauge fields for $r \to \infty$ can be obtained
from the generic formula 
\be
A_\mu =- \frac{i}{g}  U^\infty \partial_\mu U^{\dagger\infty}~~,
\ee
where $A_\mu$ represents the gauge field, $g$ is the gauge coupling
constant and $U^\infty$ is given by Eq.~(\ref{uinfp}). Accordingly,
for $p=1/2$ 
\be
W^3_{R\theta}(\infty,\theta) = {T^3_R\over 2rg},~ 
B_\theta(\infty,\theta) = {X \over 2rg'} ~~ , \label{gauge}
\ee
where $g$ and $g'$ are the gauge couplings. At finite values of the
radial coordinates $r$, Eq.~(\ref{gauge}) should be replaced by
\be
W^3_{R\theta}(r,\theta) = {T^3_R\over 2rg}\Big(1-h_R(r)\Big),~ 
B_\theta(r,\theta) = {X \over 2rg'}\Big(1-h_B(r)\Big) ~~. \label{rgauge}
\ee
The real functions $h_i(r)$ satisfy the following boundary
conditions: $h_i(0)=1$ and $h_i(r) \to 0$ as $ r \to \infty$, $i=R, B$. 

After subsequent symmetry breaking, the above mapping
$U^\infty(\theta)$  (i.e. the map of Eq.~(\ref{uinfp}) with $p=1/2$)
does not suffice to signal the nontrivial sector. The low temperature
vevs of the $(1,0,1)$ field $\Delta_L$ and the $(1/2,1/2,0)$ 
field $\phi$ are respectively (see Eq.~(\ref{vevs}))
\be
\Delta_{L} = \pmatrix{ 0 & 0 \cr v_{L} & 0} ~, ~~
\phi = \pmatrix{\kappa & 0 \cr  0 & \tilde\kappa} ~~.
\ee
These fields are not invariant under the action of
$U^\infty(2\pi)$. However, one may think of this curve
$U^\infty(\theta)$ as a projection to the subspace $SU(2)_R\otimes
U(1)_{B-L}$ of the more general curve
\be
{\tilde U}^\infty(\theta) = \exp\{i(T^3_L+T^3_R-X)\theta/2\}
~~. \label{gcurve} 
\ee
It can be easily shown that the above mapping ($p=1/2$), $\tilde
U^\infty(\theta)$, leaves $\Delta_R(\infty,\theta)$ to be as in
Eq.~(\ref{deltarp}) and gives $\theta$ dependence to the  $\Delta_L$ vev
as follows 
\be
\Delta_L(\infty, \theta) = \pmatrix{0 & 0 \cr e^{-i\theta}v_L & 0} ~~ .
\ee
Since the $\Phi$ vev is diagonal, it remains invariant under the action of
the mapping in Eq.~(\ref{gcurve}). The reason for the topological
stability of this sector is that $\tilde U^\infty(2\pi)$ belongs to
$(-I,-I)$ sector. The ansatze for gauge fields in the low temperature
phase can be derived from Eq.~(\ref{gcurve}). These are given by
Eq.~(\ref{rgauge}) and an additional real function $h_L(r)$  for the
$W^3{_L}$ field. 

An important conclusion of the discussion so far is that this model
predicts cosmic strings to exist at the present epoch. At earlier
epochs, their dynamics may be treated by methods that have now become 
standard \cite{vilenshel}. An important point emerges from our
analysis. The vev of the bidoublet field $\Phi$ dicatates that if
$\Delta_L$ vev lies 
in the nontrivial sector, so must the vev of $\Delta_R$. Hence no string
can exist without $SU(2)_R$ magnetic flux in its core. This is
essential for estimating the abundance of such strings at any
epoch. Further, their string tension is dominated by the scale $v_R$. 

There is an additional use of the path $U^\infty(\theta;p)$ 
identified in Eq.~(\ref{uinfp}). If we choose
$p=1/4$ rather than $1/2$, we obtain
$\Delta_R(\infty,0)=-\Delta_R(\infty,2\pi)$. We show in the next
section that such configurations can be the boundaries of domain
walls. Such domain walls separate regions with opposite signs for
the vev of $\Delta_R$.

The cosmic strings also carry fermion zero-modes. The equation
governing a fermion field $\psi$ in the background of a vortex has the 
form 
\be
i \rlap{\,\slash}{D} \psi + \frac{\delta}{\delta \bar\psi} {\mathcal
L}_Y = 0 ~~, 
\ee
where ${\mathcal L}_Y$ is given by  Eq.~(\ref{yukawa}),
$\rlap{\,\slash}{D}= \gamma^\mu D_\mu$ and the
background gauge fields have to be substituted in the covariant
derivative $D_\mu$.

The charged fermions do not couple to the $\Delta_L$, $\Delta_R$,
and since $\phi$ vev remains trivial as in
Eq.~(\ref{vevs}), we find for the quarks and the leptons
\be
\gamma^0 \partial_o \psi_F + \gamma^i (\partial_i + i \tilde
Q_F(A_{bg})_i)\psi_F - m_F \psi_F = 0 ~~, \label{zeromode}
\ee
where $F$ is a label for the fermionic species, $m_F$ is the fermionic
mass derived from coupling to $\phi$, $A_{bg}$ is the background gauge
field and $\tilde Q_F$ is the value of 
$(T_L^3+T_R^3-X)$ charge of the fermion. The presence of the 
mass term precludes the possibility of zero mode solutions at low
temperatures. Above the electroweak scale, the $\phi$ vev disappears
and $\tilde Q$ has to be replaced by $\tilde Y$. The condition for
existence of zero-energy normalizable solutions is that $|\tilde Y| >
1$ \cite{jrew}. The number of zero modes is then equal to the largest
integer less that $|\tilde Y|$. The $\tilde Y$ charge of the left
handed leptons and the left handed quarks is $1/2$ and 
$-1/6$ respectively . For the right handed fermions 
$u_R$, $d_R$ and $e_R$ it is $1/3$, $-2/3$ and $0$ respectively.
All of these fermions do not posses zero-energy modes coupled to
cosmic strings.  

For the left and right handed neutrinos, $N_L$ and $N_R$, the
equations of motion are 
\begin{eqnarray}
\partial_0 N_L + \tau^i \left(\partial_i + i \tilde Q_L
(A_{bg})_I \right) N_L -f v_L e^{-i \theta} N_L =0~,
\nonumber \\
\partial_0 N_R + \tau^i \left(\partial_i + i \tilde Q_R
(A_{bg})_I \right) N_R -f v_R e^{-i \theta} N_R =0
~~, 
\end{eqnarray}
where $f$ was introduced in Eq.~(\ref{yukawa}) and, $v_L$ and
$v_R$ are the vevs of $\Delta_L$ and $\Delta_R$ fields
respectively. The existence and number of zero-modes is determined by
the $\theta$-dependent scalar coupling. If $\theta$ winds $m$ times
around the unit circle, there are $m$ zero modes. Accordingly, both
the neutrinos posses solitary zero modes. At higher temperatures,  
$\langle L \rangle =0$ and the existence of the $N_L$ zero-modes is
determined by the $\tilde Y$ charge. This howerver is $1/2$ and no
zero modes result.

\end{section}
%
%
\begin{section}*{IV. DOMAIN WALLS }
 The minimal left-right symmetric model possesses more than one kind
of domain wall (DW) solutions. A solution for which the nonzero
component of $\Delta_R$ is proportional to $\tanh(ax), x=0$ being the
plane of the DW, is readily obtained. This solution has
$\Delta_L=\tilde\phi=\phi=0$ and is therefore trivial. A different,
nontrivial solution also exists, as can be seen by considering the
full scalar potential
$V(\Delta_L,\Delta_R,\phi,\tilde\phi)$ (see Eq.~(\ref{potential})). 

 We assume that the ansatz functions 
$L(x)$, $R(x)$, $f(x)$ and $\tilde f(x)$ are the
nonzero components of $\Delta_L$, $\Delta_R$, $\phi$ and $\tilde\phi$
respectively. By minimizing the energy, the equations of motion
governing the wall configurations are
\be
L^{''}(x)=\frac{\partial V}{\partial L(x)} \, , \quad
R^{''}(x)=\frac{\partial V}{\partial R(x)} \, , \quad
f^{''}(x)=\frac{\partial V}{\partial f(x)} \, , \quad
\tilde f^{''}(x)=\frac{\partial V}{\partial \tilde f(x)} ~~,
\ee
where the prime means the derivative with respect to $x$. The boundary
conditions, as $x \to \pm\infty$, are \begin{eqnarray}
L(x) \to \pm v_L  \nonumber \, , \\ 
R(x) \to \pm v_R  \nonumber \, , \\ 
f(x) \to \pm \kappa \nonumber \, , \\
\tilde f(x) \to \pm \tilde\kappa ~~.  \label{vacuum}
\end{eqnarray}
\vskip 0.6 cm
{\bf(A) Left-right domain wall solutions}
\vskip 0.2cm
 At tree level the Lagrangian is symmetric under the exchange
$\Delta_L \leftrightarrow\Delta_R$, reflecting the hypothesis 
of left-right symmetry. The vacuum values for these
two Higgs fields are $v_L$ and $v_R$ (see Eqs.~(\ref{vevs}) and
(\ref{vacuum})). It can be shown \cite{mohapbook} that the triplet
part of the potential, defined in Eq.~(\ref{vdelta}), takes the form 
\begin{eqnarray}
V(\Delta_L, \Delta_R) = -\mu^2(\Delta_L^2 + \Delta_R^2)
+ (\rho_1+\rho_2)(\Delta_L^4+\Delta_R^4) + \rho_3\Delta_L^2\Delta_R^2
~~. \label{treepot}  
\end{eqnarray}
 Upon parameterizing $\Delta_L=v\sin\alpha$ and $\Delta_R=v\cos\alpha$,
Eq.~(\ref{treepot}) reads
\be
V(v,\alpha)= -\mu^2 v^2 + v^4\Big(\rho_1+\rho_2+\frac{1}{4}\beta
\sin^2(2\alpha)\Big) ~~, \label{V(v,alpha)}
\ee
where $\beta=\rho_3 -2(\rho_1+\rho_2)$.

 The points $(v, \alpha)=(v_0, 0)$ and 
$(v_0, \pi/2)$ with $v_0=\sqrt{\mu^2/2(\rho_1+\rho_2)}$
are the minima, and 
$\Big(\sqrt{2\mu^2/(\rho_3+2(\rho_1+\rho_2))}, \pi/4\Big)$
a saddle point, provided $\beta>0$. 
Electroweak phenomenology dictates that the latter condition
be valid.

It is reasonable to assume that the
effective potential continues to enjoy the above discrete
symmetry, since the same loop corrections enter for both the
fields. This means the symmetry is broken spontaneously
at the left-right breaking scale, providing requisite topological
conditions  
for the existence of domain walls. As the universe cools from the
left-right symmetric phase, there are causally
disconnected regions that select either $\alpha=0$ or
$\alpha=\pi/2$. Thus the vevs are functions of position 
and the two kinds of regions are separated by domain walls. 

 We further define
\be
\sigma(x) = \sqrt{R(x)^2+L(x)^2}  ~, \quad
\xi(x) = \tan^{-1}{L(x) \over R(x)} ~~. 
\ee
 Then the equations of motion take the form
\begin{eqnarray}
\frac{d^2\sigma}{dx^2} & = & -2\mu^2\sigma +
4\sigma^3\Big[(\rho_1+\rho_2)+\frac{1}{4}\beta 
\sin^22\xi\Big] - \sigma \Big(\frac{d\xi}{dx}\Big)^2 ~~, \cr
&&\cr
\frac{d}{dx}\Big(\sigma^2\frac{d\xi}{dx}\Big)  & = & 
{1\over2}\sigma^4 \beta \sin4\xi ~~.
\label{LRdweqns}
\end{eqnarray}
 The boundary conditions appropriate to the DW are
\begin{eqnarray}
\sigma(x) \to v_0  \quad {\rm as} ~ x \to \pm\infty ~~, \nonumber \\
\xi(x) \to 0  \quad {\rm as} ~ x \to -\infty ~~, \nonumber \\
\xi(x) \to \pi/2 \quad {\rm as} ~ x \to +\infty  ~~, \label{bc1}
\end{eqnarray}
or alternatively,
\begin{eqnarray}
R(x) \to v_0 ~, ~~ L(x) \to 0 \quad {\rm as} ~ x \to -\infty ~~, \nonumber \\
R(x) \to 0 ~, ~~ L(x) \to v_0 \quad {\rm as} ~ x \to +\infty ~~ . \label{bc2}
\end{eqnarray}
In particular, for 
$\rho_3 = 6 (\rho_1 + \rho_2)$, one finds an exact solution
\be
R(x) = \frac{v_0}{2} \Big[ 1 - \tanh(\mu x)\Big] ~, ~~
L(x) = \frac{v_0}{2} \Big[ 1 + \tanh(\mu x)\Big] ~~ .
\ee
In terms of $\sigma(x)$ and $\xi(x)$ the
exact solution is
\be
\sigma(x) = \frac{v_0}{\sqrt 2} \sqrt{1+\tanh^2(\mu x)} ~, ~~ 
\xi(x)=\tan^{-1}\Big[\frac{1+\tanh(\mu x)}{1-\tanh(\mu x)}\Big]   ~~ .
\ee
If $\beta$ is very small, then we get the approximate solution 
\be
\sigma^2(x)=\frac{\mu^2}{2(\rho_1+\rho_2)} ~, ~~
\xi(x) =\tan^{-1}\Big[\exp\{\mu x\sqrt{2 \beta/(\rho_1+\rho_2)}\}\Big] 
~~. \label{approx} 
\ee
 We have found a numerical solution for the domain wall configurations
$\sigma(x)$ and $\xi(x)$ by minimizing 
the energy for different values of the parameters in the potential
Eq.~(\ref{V(v,alpha)}). Figure 1 shows the numerical result for the
domain walls for
$\rho_1+\rho_2=0.1$, $\rho_3=0.9$, $\mu^2=1$ and $\beta=0.7$. Figure 2 
shows the results for
$\rho_1+\rho_2=0.05$, $\rho_3=0.6$, $\mu^2=1$ and $\beta=0.5$. Figure 3 
shows the result for
$\rho_1+\rho_2=0.2$, $\rho_3=0.5$, $\mu^2=1$ and $\beta=0.1$. As we
can see from the figures, as $\beta$ decreases, the solution
approaches the approximate solution given by Eq.~(\ref{approx}). These
results are confirmed by solving Eqs.~(\ref{LRdweqns}) numerically.

 At the electroweak scale, the effective potential does not respect
left-right symmetry due to the nature of the $\phi$ self coupling.
One finds that $v_Lv_R \sim \kappa^2$. Upon choosing
$\kappa \sim v_{EW}$ with $v_{EW}$ 
denoting the electroweak scale, $v_L$ is driven to be tiny.
The $Z_2$ guaranteeing the topological stability of the walls
now disappears. Energy minimization requires that the walls 
disintegrate. 

 There is a possibility that the left-right symmetry is not exact due to
effects of a higher unification scale. In that case, the R breaking
minimum should be energetically preferred by small amounts before 
the electroweak phase transition. This will cause the
domain walls  to move around till the regions
with the L breaking false vacuum have been converted to the true
vacuum. Some fraction of the walls would then disappear before the
electroweak scale is reached. The fate of the surviving walls
is the same as that discussed in the previous paragraph.
Further consequences are discussed in the next section.

%
\vskip 0.6cm
{\bf(B) Domain wall solutions with $\phi$ condensate}
\vskip 0.2cm
 In order to have the observed near-maximal parity violation at low 
energies, we must have $\kappa \ll v_R$. Also, to avoid fine tuning in 
the potential we must have $\tilde\kappa=0$. But $v_L \ll
\tilde\kappa$, so we shall set $v_L=0$ \cite{Gunion}. So we are left
with only two fields $\Delta_R$ and $\phi$. The field $\Delta_R$ admit
a domain wall solution where the field $\phi$ develops a condensate in
the core of the domain wall. The potential in Eq.~(\ref{potential}) is
simplified to
\be
V(\phi,\Delta) = \lambda C^4 \phi^4 -\mu^2_\kappa C^2 \phi^2 + \rho
\Delta_R^4 - \mu^2 \Delta_R^2 + \alpha C^2 \phi^2 \Delta^2_R ~~ ,
\label{potphikappa}
\ee
where $\lambda=\lambda_1 +\lambda'_1$,~
$\mu^2_\kappa=\mu^2_{11}+\mu^2_{22}$~, $\rho=\rho_1+\rho_2$,~
$\alpha=\alpha_{11}+\alpha_{22}+\beta_{11}$ and
$C={\kappa}/{v_R}$ (see Eq.~(\ref{potential})). Since the potential
of Eq.~(\ref{potphikappa}) is 
invariant under the discrete symmetry $\Delta_R \leftrightarrow
-\Delta_R$, domain walls are formed when this symmetry is 
spontaneously broken by field $\Delta_R$ acquiring a nonzero vacuum
expectation value $v_R$. At the electroweak scale the $\phi$ field
acquires a vev $\kappa$ and forms a condensate inside the domain wall.  
We use, as before, the ansatz functions $R(x)$ and $f(x)$ 
for the nonzero components of the fields $\Delta_R$ and $\phi$
respectively, where the boundary condition is $R(x) \to \pm v_R$ 
as $ x \to \pm \infty$. We choose the origin of $x$ such that 
$R(0)=0$. We have minimized the energy for different values of the
parameters in the potential Eq.~(\ref{potphikappa}). Figure 4 shows
the numerical results for the DW profile for $\rho=0.5$, $\lambda=0.01$,
$\alpha=\mu^2_\kappa=0.4$, $v_R=1.0$ and $C=0.01$ while Fig. 5 shows
the results for $\rho=0.3$, $\lambda=0.1$,
$\alpha=\mu^2_\kappa=0.3$, $v_R=1.28$ and $C=0.01$. Finally, Fig. 6
shows the results for $\rho=1.0$, $\lambda=0.01$,
$\alpha=\mu^2_\kappa=0.0.01$, $v_R=0.7$ and $C=0.01$. 

 It is interesting to consider the ultimate fate of these domain
walls. As we have shown in Sec. III, if only $SU(2)_R$ is 
broken, then a topologically  unstable cosmic string may be formed
with $\Delta_R(\infty,0)=-\Delta_R(\infty,2\pi)$. Since the DW has the 
boundary condition $R(x) \to \pm v_R (= \pm \Delta_R(\infty,0))$ as 
$x \to \pm \infty$, these 
unstable strings will be the boundary of the DW. The dynamics of the
cosmological netwotks of string-bounded walls has been studied
\cite{kiblazshafi}. The walls eventually shrink via surface tension,
string intercommutation and nucleation of new string loops. Thus they
never dominate the energy density of the universe, and can have
interesting cosmological effects while they last.

\end{section}
%
%
\begin{section}*{V. CONCLUSIONS}
From the point of view of a predictable baryogenesis,
the left-right symmetric model enjoys the advantage that the
primordial value of the 
$B-L$ number is naturally zero, being the value of an
Abelian gauge charge. The topological defects studied here can play a 
significant role in baryogenesis through leptogenesis.

 In the context of left-right symmetric models, mechanisms for 
electroweak baryogenesis that do not rely on topological defects have 
been investigated recently \cite{mohzha,freetal}.
It has been shown that the parameters in the 
potential, for the minimal model considered here,
require unnatural fine tuning to provide sufficient 
CP violation to expalin the observed baryon asymmetry.

 It has been shown recently that spontaneous CP violation can occur
in the minimal left-right symmetric model considered here
\cite{barenboim}. However, baryogenesis with only spontaneous breakdown 
of CP presents severe cosmological problems, due to the formation of
domain walls as a result of the breaking of a discrete
symmetry. Moreover, in order to generate
baryon asymmetry, the scale of the spontaneous CP violation and the
scale at which the baryogenesis takes place must be different
\cite{reina}; otherwise, an equal amount of matter and anti-matter is
generated. In the minimal left-right symmetric model with spontaneous 
CP violation, both scales coincide and therefore electroweak
baryogenesis is not feasible.

 Defect-mediated leptogenesis mechanisms also need enhanced
CP violation. For the present purpose we note that the $\sigma$ 
field \cite{mohzha} does not alter the topological considerations
presented above since it is a gauge singlet and
its main function is to bias the potential of the field $\phi$. The
coupling of $\sigma$ to both $\Delta_L$ and $\Delta_R$ 
may be assumed to be identical due to left-right symmetry. Then the
domain walls present very interesting prospects. 
Their interaction with other particles in the pre-electroweak 
scale plasma can result in leptogenesis.
A model-independent possibility of this kind was 
considered in \cite{rbetal2}. More specific considerations also
appear in \cite{mencoo} and \cite{lewrio}. 
It is likely that the model is descended from a grand unified theory.
For this or for some other reason there may be a small
asymmetry between the L-preferring and R-preferring minima even 
above the electroweak scale. If the energy density difference 
is suppressed by powers of the GUT mass, the walls 
are still expected to be present long enough to bring about
requisite leptogenesis. 

 The case of exact left-right symmetry leads to domain walls that
are stable above the electroweak symmetry
breaking scale. In this case the regions trapped in ${\vev \Delta_L} \sim
v_R$ vacuum will become suddenly destabilized as the $\phi$ acquires
a vev. The destabilization can generate large amounts of entropy
and the domains should reheat to some temperature $T_H$ greater
than $v_{EW}$ but less than $v_R$. The possibility for baryogenesis
from situations with large departure from thermal equilibrium
was considered by Weinberg\cite{Weinberg}. It was argued
that in such situations the asymmetry generated should be
determined by
the ratio of time constants governing baryon number violation
and entropy generation respectively. In the present case we
expect leptogenesis from the degeneration of $\vev{\Delta_L}$ 
due to the Majorana-like Yukawa coupling mentioned in Sec. II.
The generated lepton 
asymmetry can then convert to baryon asymmetry through the 
electroweak anomaly. This possibility will be studied separately.

We have also shown that model admits DW
solutions if we impose the phenomenological hierarchy $v_L \ll 
\tilde\kappa$, $\kappa  \ll
v_R$, and avoid fine tuning in the
potential. The fields in this case are $\Delta_R$ and $\phi$. 
Domain walls are formed when $\Delta_R$ field acquires a  
 non-zero vev.  At the electroweak scale the
$\phi$ field acquires a vev $\kappa$ and forms a condensate inside the 
domain wall. Since these domain walls formed at the same scale as the 
unstable cosmic strings, the unstable strings will be the boundary of
the DW. These DW solutions are unstable: they will shrink and disappear.

The cosmic strings demonstrated above can play several 
nontrivial roles in the early universe. They can provide sites
for electroweak baryogenesis as proposed in \cite{rbetal2}.
It has also been proposed that the fermion zero modes they 
possess can result in leptogenesis
\cite{rj1}.  Equally interesting is the 
process of disintegration of the unstable strings below
the electroweak scale. The decay should proceed by appearance
of gaps in the string length with formation of monopoles
at the ends of the resulting segments. The free segments then
shrink, realizing the scenario of \cite{rbetal3}. 

The left-right symmetric model considered here provides a
concrete setting for all of the above scenarios.
Several new features that have been demonstrated can alter
the scenarios qualitatively and merit further study. 

\end{section}
\section*{Acknowledgement}
This work was started at the 5th workshop on High Energy 
Physics Phenomenology (WHEPP-5) which was held in IUCAA, Pune,
India and supported by S. N. Bose National Centre for Basic Science, 
Calcutta, India and Tata Institute of Fundamental Research, Mumbai,
India.  H.W. thanks the University Grants 
Commission, New Delhi, for financial support. The work of U.A.Y. is
supported in part by a Department of Sceince and Technology grant.

\newpage

\begin{section}*{Figure Caption}
\begin{enumerate}

\item[] FIG. 1. Domain wall solutions for
$\rho_1+\rho_2=0.1$, $\rho_3=0.9$, $\mu^2=1$ and $\beta=0.7$. 
The solid line is the $\xi(x)$ field while the dashed line 
is the $\sigma(x)$ field. 

\item[] FIG. 2. Domain wall solutions for 
$\rho_1+\rho_2=0.05$, $\rho_3=0.6$, $\mu^2=1$ and $\beta=0.5$. 
The solid line is the $\xi(x)$ field while the dashed line 
is the $\sigma(x)$ field. 

\item[] FIG. 3. Domain wall solutions for
$\rho_1+\rho_2=0.2$, $\rho_3=0.5$, $\mu^2=1$ and $\beta=0.1$. 
The solid line is the $\xi(x)$ field while the dashed line is 
the $\sigma(x)$ field. 

\item[] FIG. 4. Domain wall solutions for $\rho=0.5$, $\lambda=0.01$,
$\alpha=\mu^2_\kappa=0.4$, $v_R=1.0$ and $C=0.01$. 
The solid line is the $R(x)$ field while the dashed line is 
the $f(x)$ field. 

\item[] FIG. 5. Domain wall solutions for $\rho=0.3$, $\lambda=0.1$,
$\alpha=\mu^2_\kappa=0.3$, $v_R=1.28$ and $C=0.01$. 
The solid line is the $R(x)$ field while the dashed line is 
the $f(x)$ field. 

\item[] FIG. 6.  Domain wall solutions for $\rho=1.0$, $\lambda=0.01$,
$\alpha=\mu^2_\kappa=0.01$, $v_R=0.7$ and $C=0.01$. 
The solid line is the $R(x)$ field while the dashed line is 
the $f(x)$ field. 

\end{enumerate}
\end{section}

\newpage
\begin{figure}[ht]
\vskip 15truecm

\includegraphics{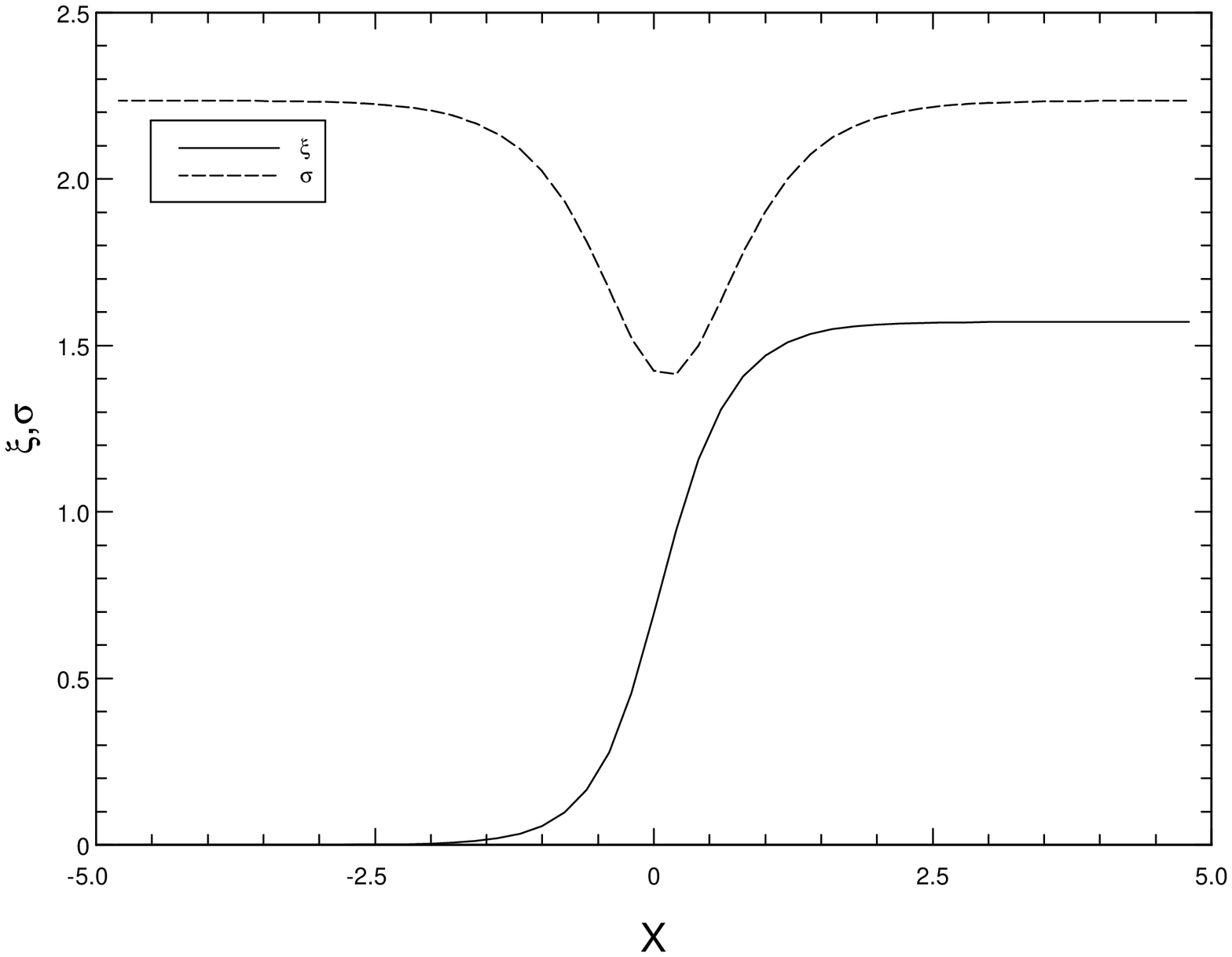} 

\caption{}
\end{figure}
\newpage
\begin{figure}[ht]
\vskip 15truecm
 \includegraphics{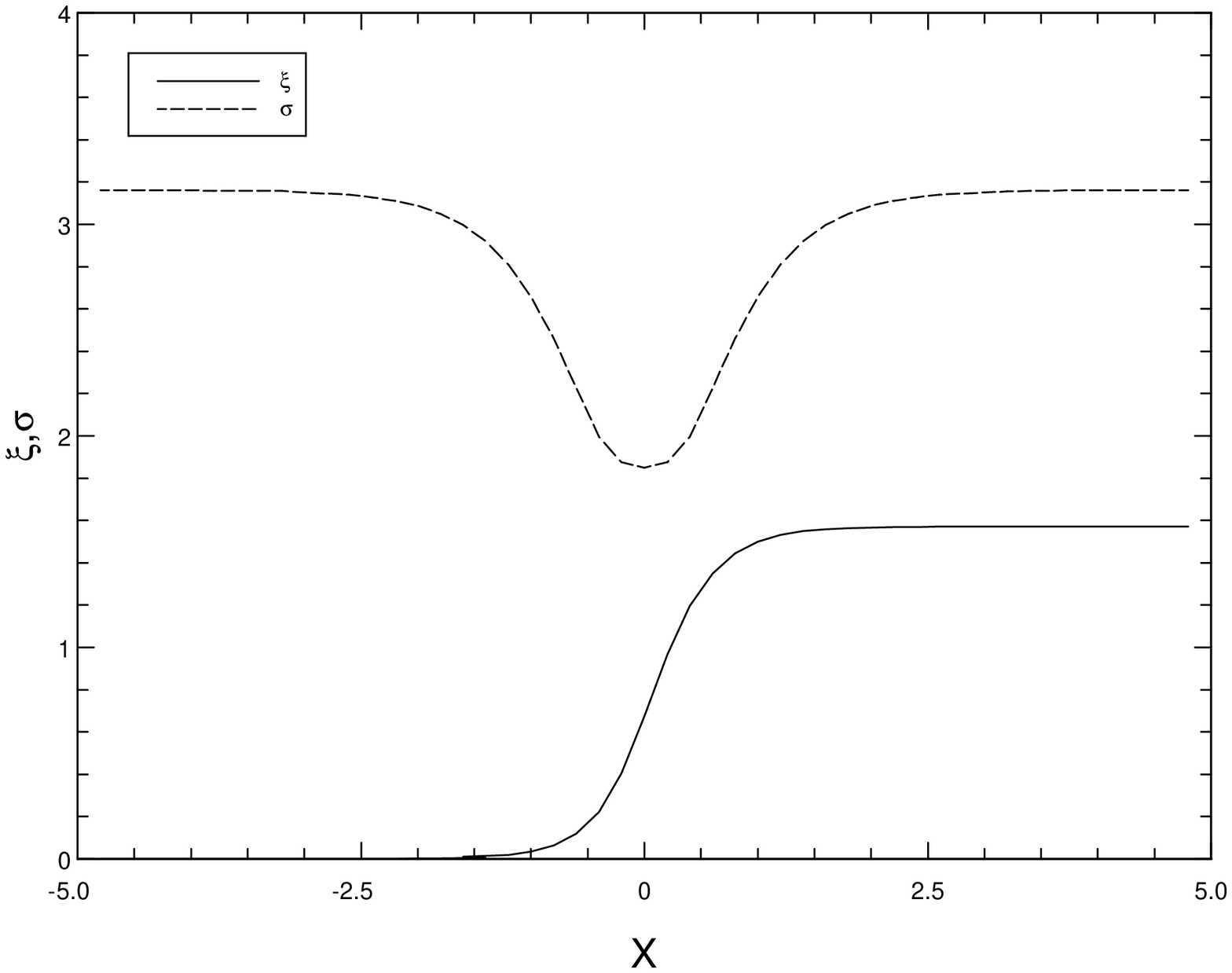} 
\caption{}
\end{figure}
\newpage
\begin{figure}[ht]
\vskip 15truecm
 \includegraphics{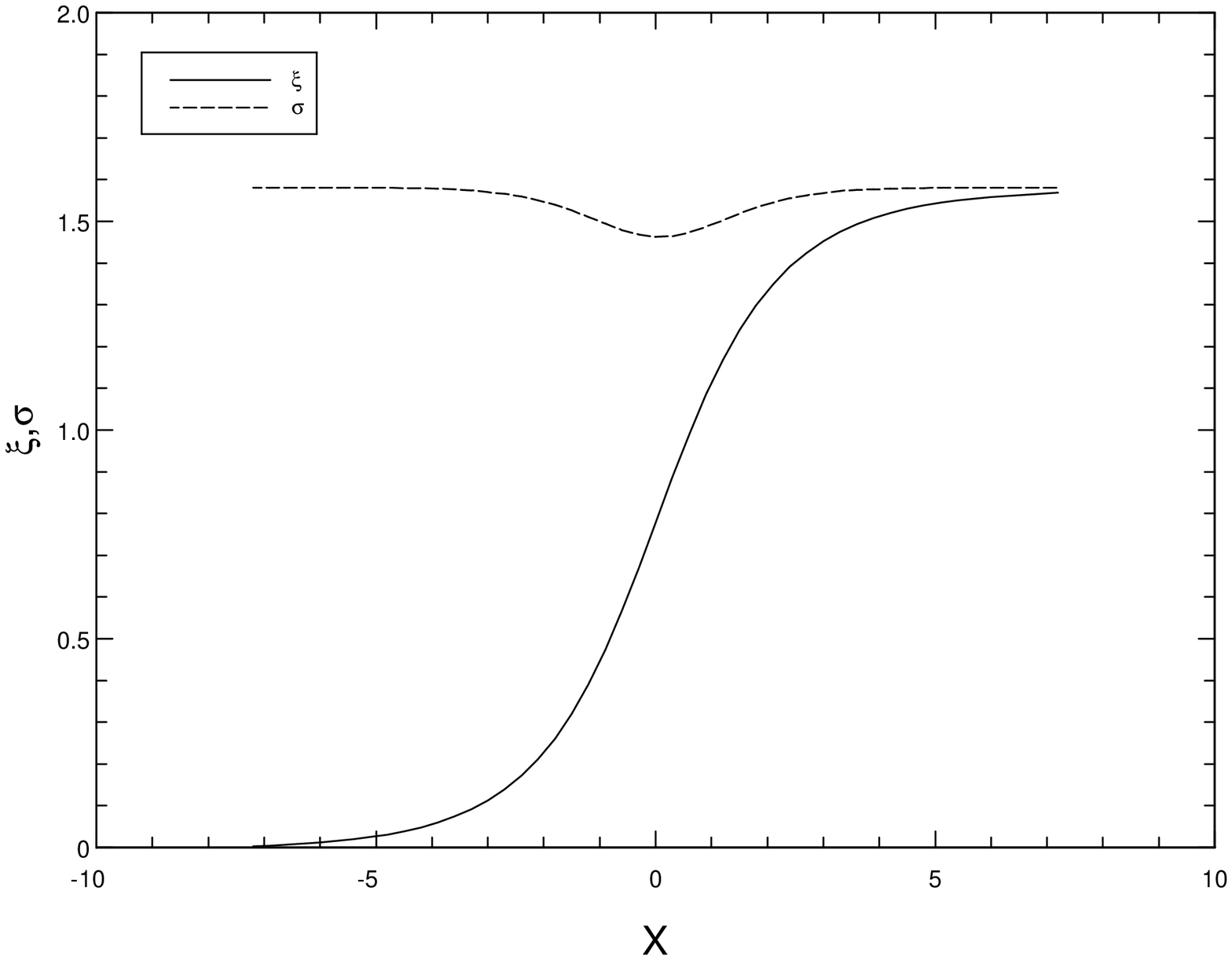} 
\caption{}

\end{figure}
\newpage
\begin{figure}[ht]
\vskip 15truecm
 \includegraphics{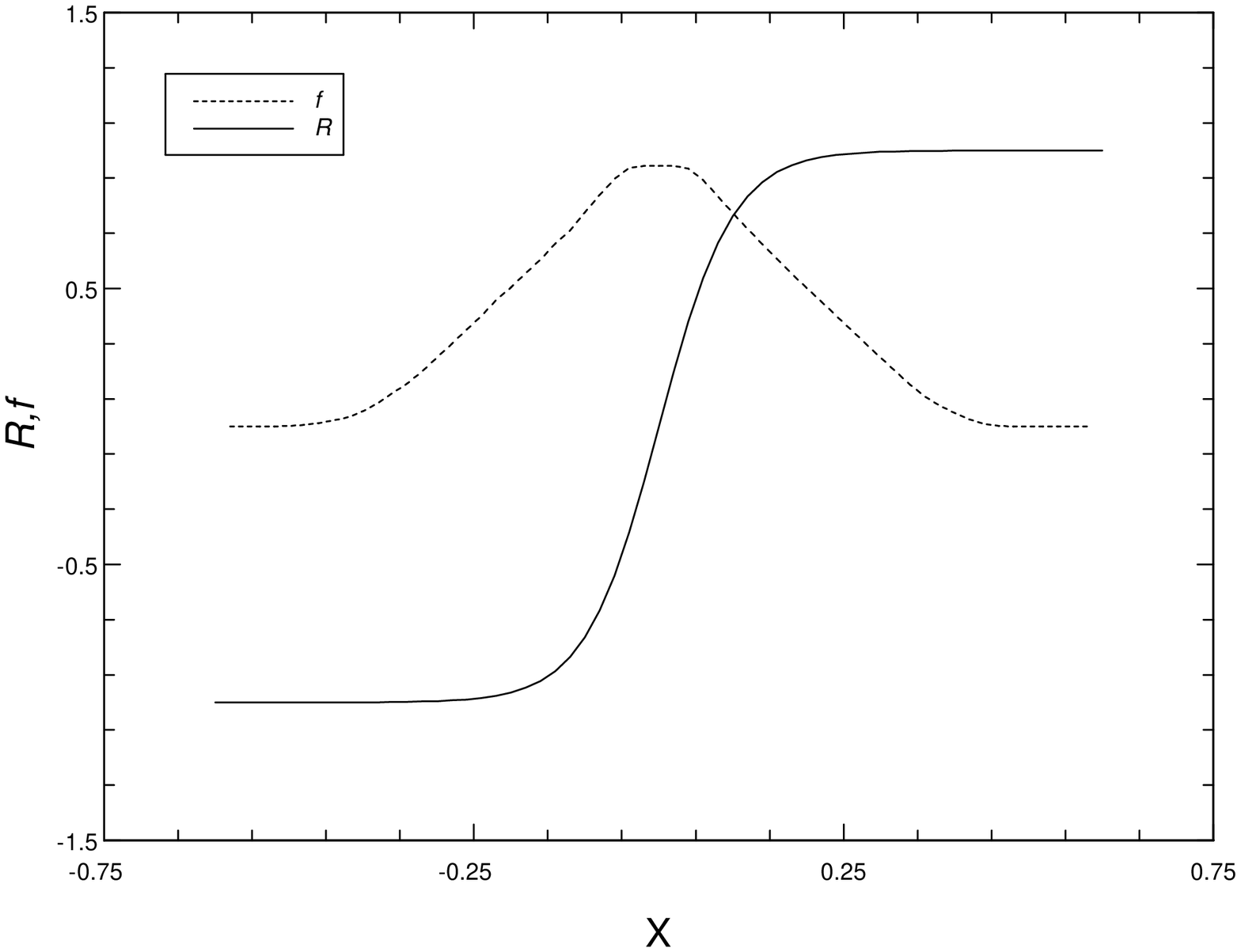} 
\caption{}

\end{figure}

\newpage
\begin{figure}[ht]
\vskip 15truecm
 \includegraphics{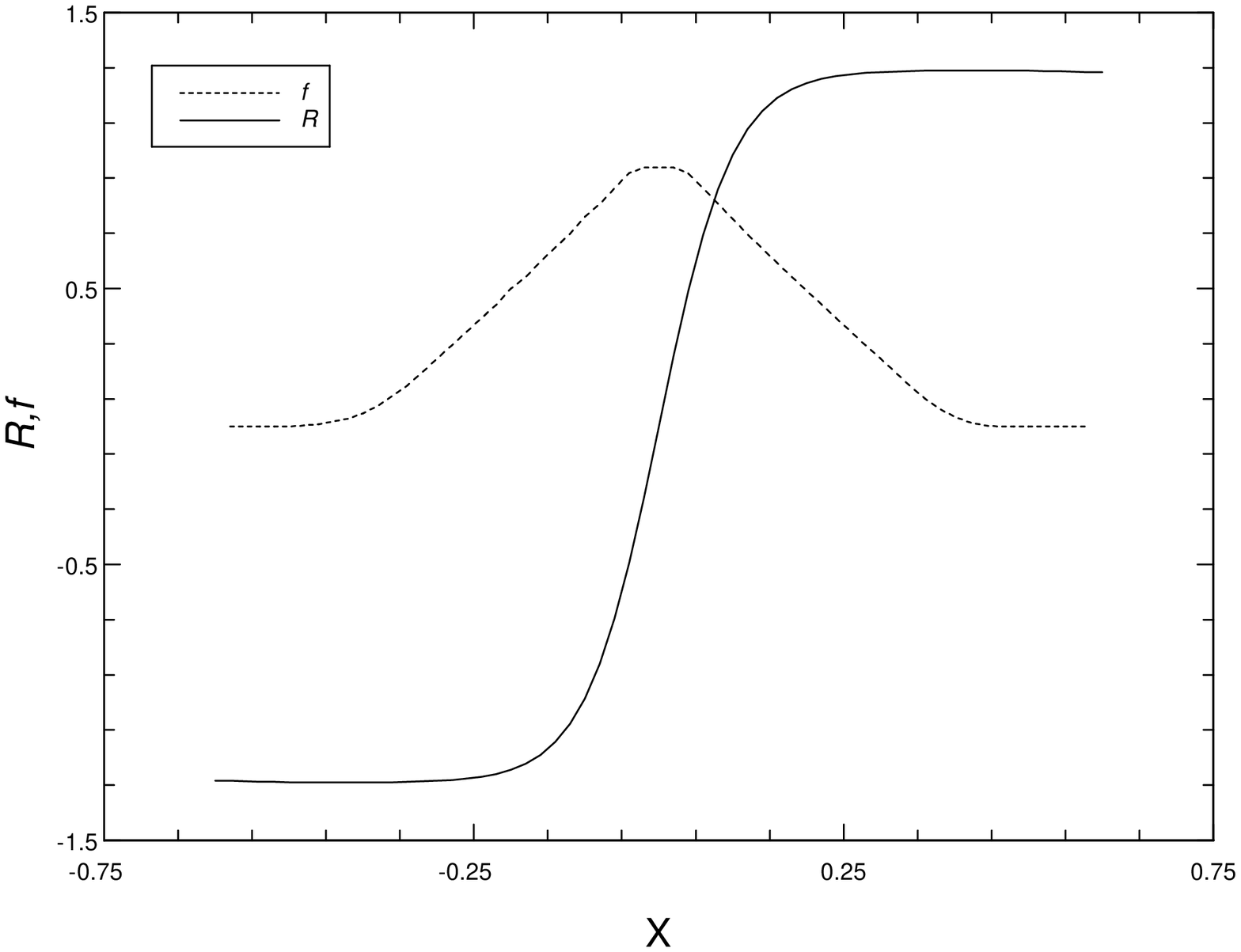} 
\caption{}

\end{figure}
\newpage
\begin{figure}[ht]
\vskip 15truecm
 \includegraphics{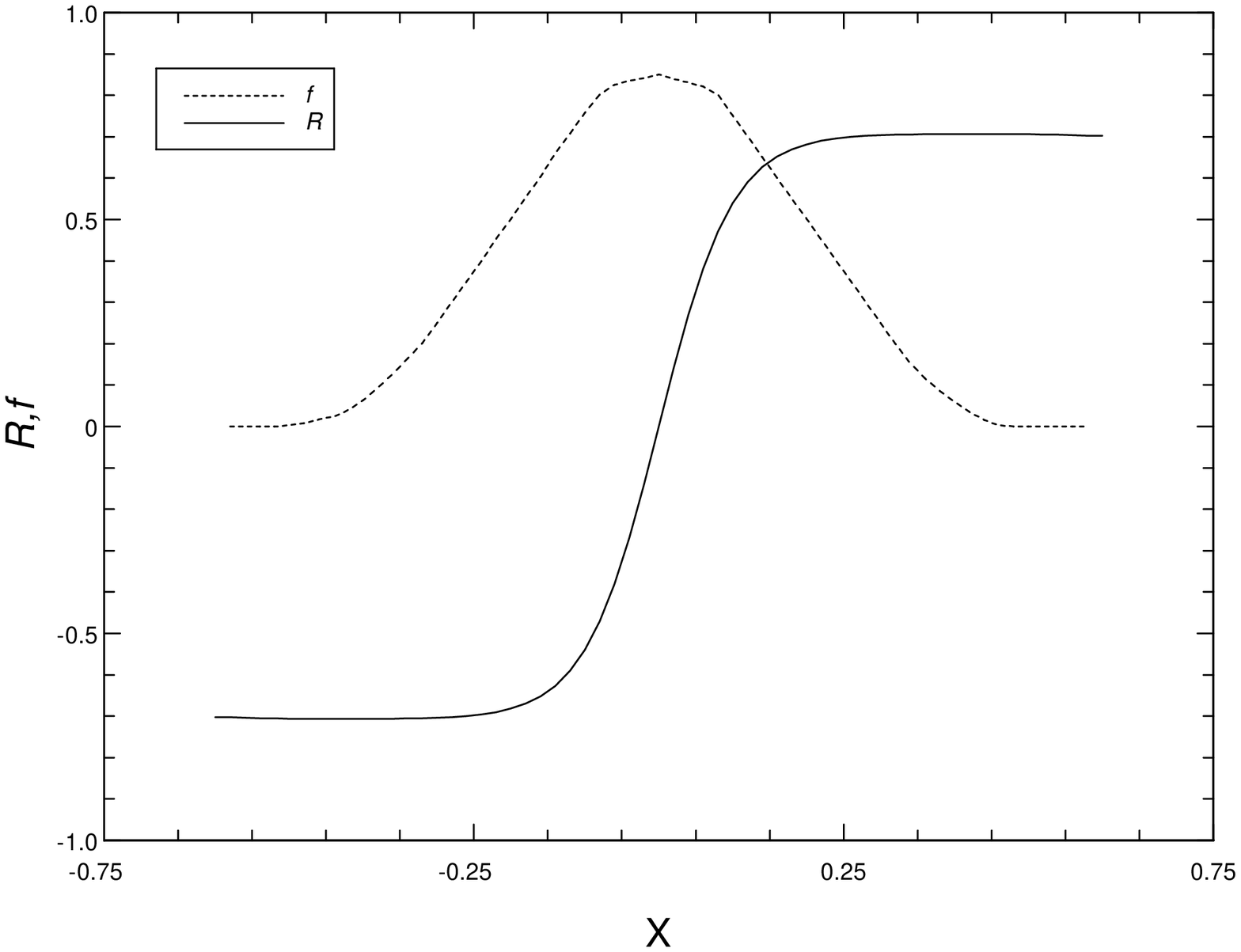} 
\caption{}

\end{figure}


\begin{thebibliography}{99}
\bibitem{footnote} We note that this reasoning does not apply to the
standard model Higgs sector 
because it is a doublet and $-I$ does not belong to the stability
group of its vev.
\end{thebibliography}
\end{document}